\documentclass[journal]{IEEEtran}
\usepackage{amsmath}
\usepackage{cite}
\usepackage{ulem}

\ifCLASSINFOpdf
\usepackage[pdftex]{graphicx}
\else
\fi

\usepackage{color}
\usepackage{booktabs}
\usepackage{threeparttable}
\begin{document}

\title{Detection of Alpha Particles and Low Energy Gamma Rays by Thermo-bonded Micromegas in Xenon Gas }

\author{Yuehuan~Wei,
Liang~Guan,
Zhiyong Zhang,
Qing~Lin,
Xiaolian~Wang,
Kaixuan~Ni
and Tianchi~Zhao~\IEEEmembership{Member,~IEEE} %
\thanks{Manuscript received October 20, 2012; revised February 17, 2013; revised April 29, 2013. 
This work is supported by National Science Foundation of China (Grant No.: 11055003, 11075156 and 11175117),
and Science and Technology Commission of Shanghai Municipality (Grant No.: 11PJ1405300) and US Department of Energy and National Science Foundation.}
\thanks{Yuehuan Wei, Qing Lin and Kaixuan Ni are with the INPAC, Department of Physics and Shanghai Key Laboratory for Particle Physics and Cosmology,
Shanghai Jiao Tong University, 800 Dongchuan Road, Shanghai, 200240, China.}
\thanks{Liang Guan, Zhiyong Zhang and Xiaolian Wang are with State Key Laboratory of Particle Detection and Electronics,
University of Science and Technology of China, 96 JinZhai Road, HeFei,230026, China.
(Corresponding author: Liang Guan, E-mail:lguan@cern.ch)}
\thanks{Tianchi Zhao is with the Department of Physics, University of Washington, Seattle, WA 98195, USA.}}

\markboth{To be submitted to IEEE TRANSACTIONS ON NUCLEAR SCIENCE}%
{Shell \MakeLowercase{\textit{Wei et al.:Detection of Alpha Particles and Low Energy Gamma Rays based on Thermo-bonded Micromegas in Xenon Gas}}}

\maketitle

\begin{abstract}
\boldmath
Micromegas is a type of micro-pattern gaseous detector currently under R$\&$D for applications in rare event search experiments.
Here we report the performance of a Micromegas structure constructed with a micromesh thermo-bonded to a readout plane,
motivated by its potential application in two-phase xenon detectors for dark matter and neutrinoless double beta decay experiments.
The study is carried out in pure xenon at room temperature.
Measurements with alpha particles from the Americium-241 source showed that gas gains larger than 200 can be obtained
at xenon pressure up to 3 atm. Gamma rays down to 8~$\rm{keV}$ were observed with such a device.
\end{abstract}


\section{Introduction}
\IEEEPARstart{D}{irect} detection of dark matter and neutrinoless double beta decay searches are of great interests in both particle physics and astrophysics.
Many experimental groups use different techniques to search for the weakly interacting massive particles (WIMPs), postulated by theories beyond the standard model of particle physics.
The two-phase noble liquid is a typical technique~\cite{Figueroa_2011} due to the large mass, self-shielding ability, scalability, high scintillation and ionization yields.
Both primary scintillation light and secondary scintillation light from ionized electrons drifting from the liquid to the gas phase, from a WIMP elastic scattering on the target nucleus,
are detected and are used for energy and position reconstruction, as well as for particle identification.
A two-phase xenon detector can also be used to search for the neutrinoless double beta decay from ${}^{136}\rm{Xe}$ simultaneously.
Various methods are used to detect these signals. So far the most successful method is to use low radioactive photo-multiplier tubes (PMTs)~\cite{Aprile_2011}.

However, PMTs are expensive and the radioactivity from them are one of the dominant background in dark matter experiments. On the other hand, micro-pattern gaseous
detectors promise a low material cost, a flexible readout pattern and high radio-purity~\cite{Cebrion_2011} which may open a new way for constructing the detector
for the next generation experiments for rare event searches~\cite{Iguaz_2011,Dafni_2011,Yahlali_2011}.
The fabrication of a Micromegas usually only involves attaching a micro-mesh on an anode board, much simpler than the fabrication of a PMT.
Thus the cost compared to a PMT-like system can be greatly reduced.
In particular, Micromegas attracts a lot of interests due to its additional features,
such as the potential to reach a very low energy threshold and an excellent energy resolution~\cite{Collar_2001}.
Micromegas with a minimum gain of 100 will be able to detect the few tens of electrons produced from a 10~$\rm{keV}$ nuclear recoil event from the interaction of a
dark matter particle in liquid xenon.
An excellent energy resolution is not crucial for dark matter detection, but will benefit the simultaneous search for neutrinoless double beta decay using the Xe detector.
Micromegas can also perform precise tracking when integrated with fine readout structures ~\cite{Iguaz_2011}.

In the past few years, two main manufacturing methods,
the bulk technique~\cite{Giomataris_2006} based on the photolithography technique and the microbulk technique~\cite{Andriamonje_2010} based on the etching of thin-kapton foils,
are successfully developed for the Micromegas. The bulk technique has already been used to produce large structures with reliable performance. In the meantime,
the microbulk technique has been used to produce Micromegas structures with excellent energy resolutions for low energy X-ray
and highly ionizing charged particles~\cite{Dafni_2009}, combined with high radio-purity for its prospects in a real experiment.

Although the bulk and the microbulk techniques have shown success in manufacturing Micromegas detectors,
we would like to explore a simple and cost-saving method to fabricate the device with similar performance.
Attaching a micromesh to the anode by using a thermo-bond film is one novel way to achieve this goal.
The bulk and microbulk techniques need the standard lithographic and etching processes,
while attaching a micromesh to the anode by using a thermo-bond film is much simpler.
The thermo-bond film is a kind of specially formulated adhesive film which is commercially available and is designed
to be used for bonding metal and plastic materials.

In this paper we report the use of thermo-bond films as separators to construct Micromegas prototypes and the test results in pure xenon gas.
The prototype fabrication and experimental setup are described in Sec.~\ref{fabric}. The performances, in terms of signal shape, electron transmission,
gas gain and energy resolution, are reported in Sec.~\ref{results}. Conclusions are drawn in Sec.~\ref{concl}.

\section{Micromegas fabrication and experimental setup}
\label{fabric}

\subsection{Micromesh and thermo-bond film}
For the fabrication of Micromegas structures, a 350-lines-per-inch stainless steel woven wire mesh was
stretched at a tension of 17-19 N/cm. The choice of the woven micromesh, in view of application in cryogenic detectors, is due
to its robustness and tolerance to the boiling noble vapour as was pointed out in~\cite{Lightfoot_2005}.

The material used to separate the micromesh from the anode readout plane is a kind of heat-active thermo-bond film which has a 51~$\mu$m thick polyester core layer
sandwiched in between two dry polymer layers of the same thickness. Both the core layer and the adhesive polymers are outstanding insulators.
Such thermo-bond film is dry at room temperature and could be activated for bonding at an internal temperature of 100-120~$\rm{{}^{o}C}$ and a pressure of 7-15~N/cm$^{2}$.
It is ideally suitable for the fabrication of micromegas avalanche gaps due to its excellent bonding strength to metal and most plastic substrates.

\subsection{Prototype fabrication and experimental setup}
The avalanche gap of the Micromegas prototype was constructed on a 90-mm-diameter printed circuit board (PCB) which has a thickness of 2.5~$\rm{mm}$.
There are 25 gold-coated copper pads on the anode PCB. Each pad is 7.9~$\rm{mm}$ $\times$ 7.9~$\rm{mm}$ in size.
The pads are spaced 0.15~$\rm{mm}$ in between and form a 5 $\times$ 5 array.

The thermo-bond film was tailored into a rectangular frame, which was placed between the stretched micromesh and the PCB.
The whole set was then moved on a flat hot plate and heat was applied from the bottom of the PCB until the thermo-bond film frame melt and became adhesive.
Pressures were given by a rubber roller from the top of the micromesh and the bonding was achieved.
After cooled down to room temperature, the micromesh was firmly attached to the anode plane at a distance of about 120~$\rm{\mu m}$ away by the thermo-bond film frame.
We also tested the bonded structure at the cryogenic temperature down to 81~$\rm{K}$ for about 12 hours.
The bonding was still solid in the cryogenic temperature and after it's recovered to the room temperature.
The active area of the prototype is defined by the unsupported mesh spans. The parameters for the constructed prototype are listed in Table~\ref{yuehua.t1}.
Detailed description of the prototype construction and test results in argon and isobutane mixtures could be found in~\cite{Liang_2011}.

\begin{table}[h]
\caption{Micromegas prototype parameters}
\centering
\begin{threeparttable}[h]
\begin{tabular}{ l l l}
\toprule
Woven MicroMesh & Wire diameter & 22 $\mu$m\\
&Thickness & 42 $\mu$m \\
&Optical transparency & 50 \% \\
\\
Avalanche gap separator & Thickness & 153 $\mu$m \\
(Thermo-bond film) & Frame Width & 1 cm\\
\\
Constructed prototypes & Active area & 30 mm $\times$ 30 mm\\
&Avalanche gap & $\sim$120 p$\mu$m\\
&Drift gap & 24 mm \\
\bottomrule
\end{tabular}
\end{threeparttable}
\label{yuehua.t1}
\end{table}

\begin{figure}[h]
\begin{center}
\includegraphics[width=0.9\linewidth]{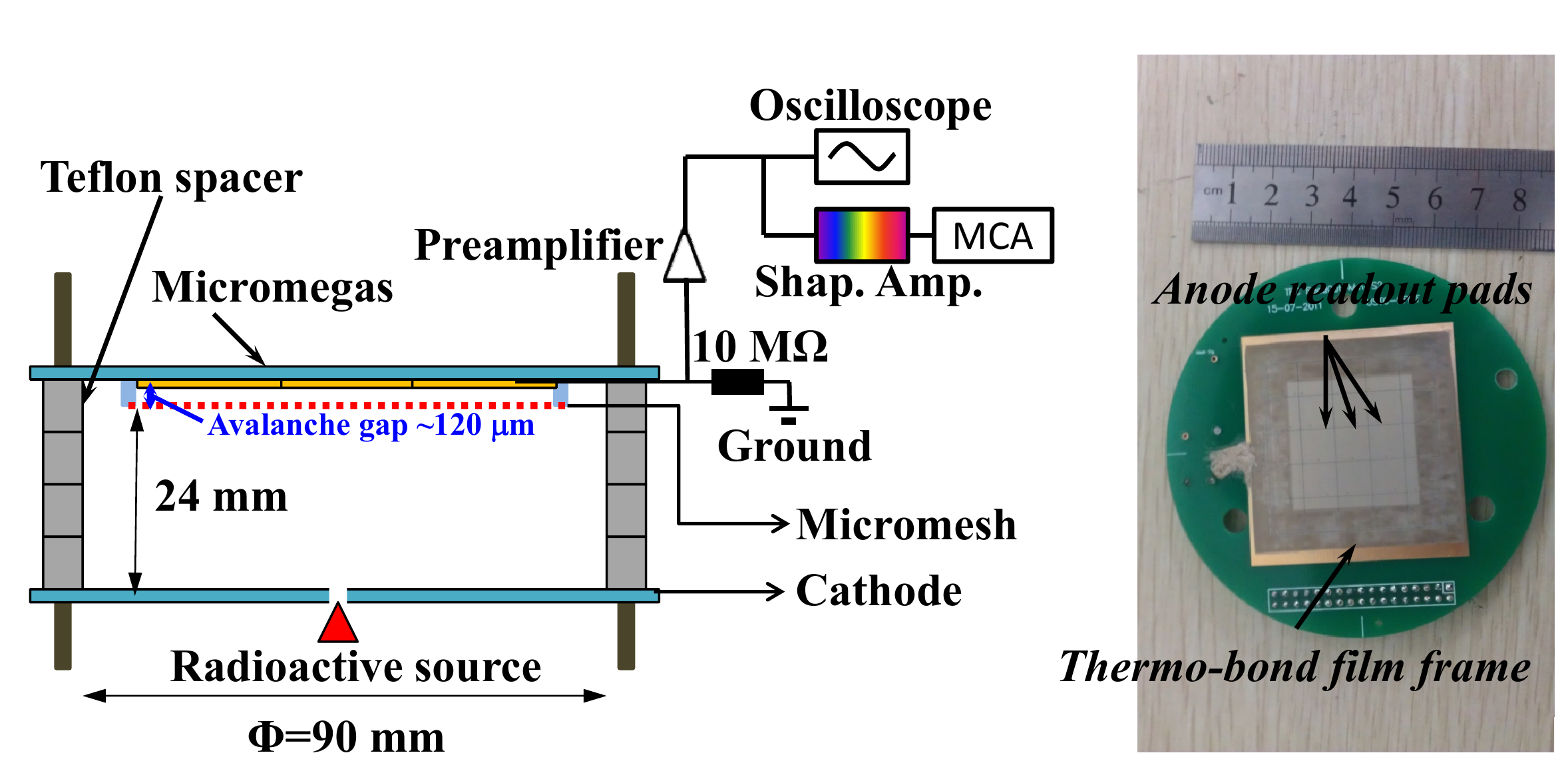}
\end{center}
\caption{A schematic view of the measurement setup (left) and a photo showing a fabricated Micromegas readout panel with the woven mesh attached to the anode
through a thermo-bond film frame (right).}
\label{yuehua1}
\end{figure}

The experimental setup for evaluating the prototype performance is illustrated in Fig.~\ref{yuehua1}.
The fabricated Micromegas as a charge gain device is fixed in the upper flange of a stainless steel vessel by three rods made of polyether ether ketone (PEEK).
A cathode made of a 0.3-mm-thick copper-coated epoxy sheet is separated by 24~$\rm{mm}$ from the micromesh using Teflon spacers.
A $^{241}$Am source with a radioactivity of 33~kBq is placed close to the cathode.
A 1~$\rm{mm}$ diameter small hole drilled on the cathode allows the penetration of alpha particles.
The distance from the radioactive source surface to the drift gap is 0.7~$\rm{mm}$.

Before the experiment, the stainless steel vessel was first pumped to a vacuum of $10^{-5}$~$\rm{mbar}$ and pure xenon gas was then filled
inside. The detector was operated in a gas sealed mode without any purification in this work.
Measurements were performed with absolute pressure from 1 atm up to 3~$\rm{atm}$ at room temperature.
For all the measurements, negative high voltages were applied on the micromesh and cathode through high voltage feed-throughs. The anode pads were grouped together and
set to the same ground through a 10~$\rm{M}\Omega$ resistor.
Signals from the anode panel were amplified by an Ortec 142AH or a ClearPulse 580 charge sensitive amplifier placed outside the vessel.
The pre-amplifier time constant is about 500~$\rm{\mu s}$ and 1~$ms$ respectively.
The amplified signals were either fed into an electronics chain containing a shaping amplifier and a multi-channel analyzer (MCA) for spectrum analysis or read out by a LeCroy Waverunner
104Xi digital oscilloscope with a 50~$\Omega$ coupling for pulse shape analysis.

\begin{figure}[h]
\begin{center}
\includegraphics[width=0.9\linewidth]{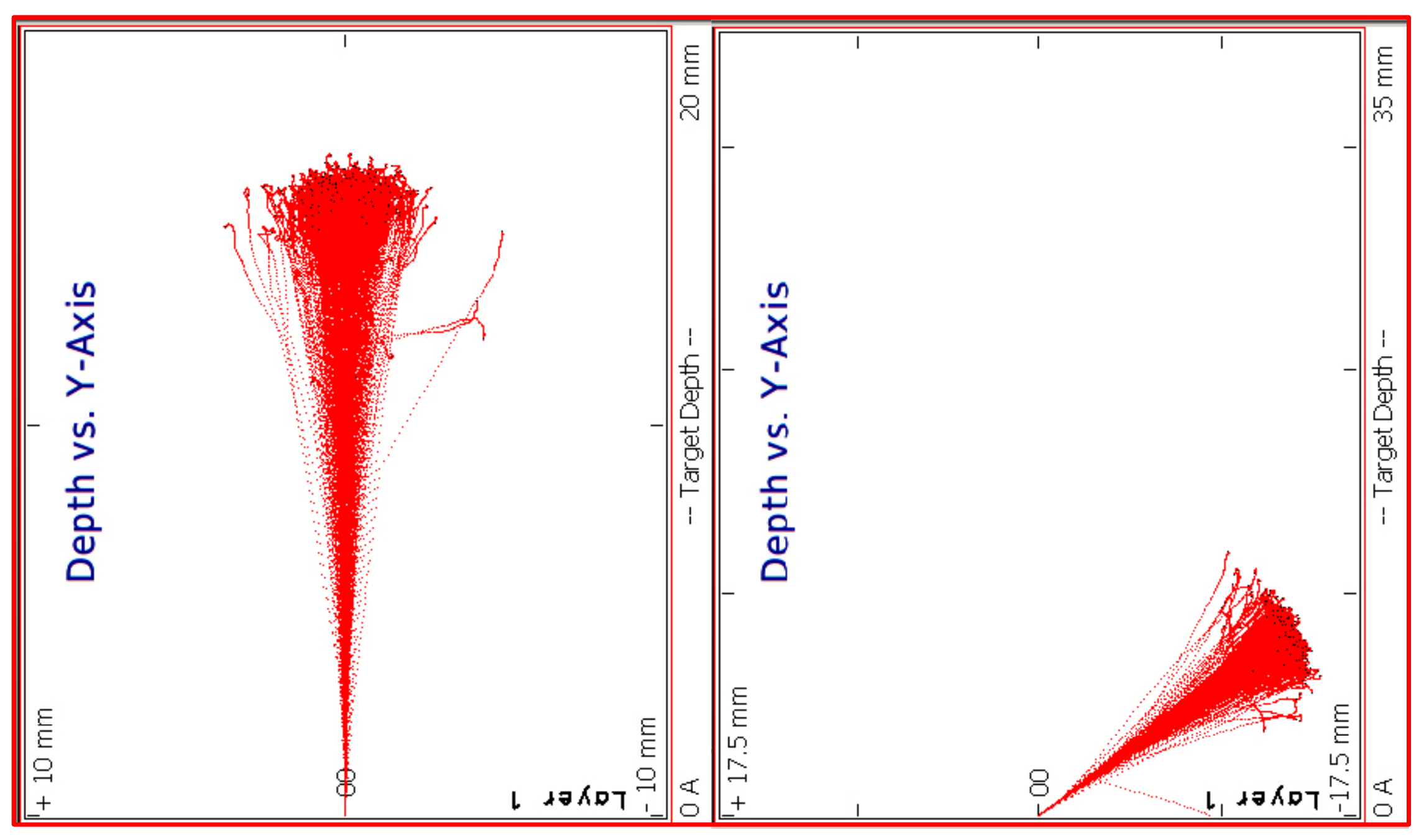}
\end{center}
\caption{4.4~$\rm{MeV}$ alpha tracks simulated by SRIM, showing the alpha tracks perpendicular to the cathode plane (left)
and the alpha tracks more parallel to the cathode plane (right). Each plot contains 500 alpha particles.}
\label{yuehua2}
\end{figure}

The 5.49 MeV alpha particles from Americium-241 have to pass through 0.5 $\mu$m thick AmO$_{2}$ substrate and
2 $\mu$m thick gold coating on source surface which leads to the loss of energy. The average energy loss of 5.49 MeV alphas in
the source is simulated using the SRIM~\cite{Ziegler_2010} package and is found to be about 1.09 MeV.
The alpha tracks either perpendicular or more parallel to the cathode plane are illustrated in Fig.~\ref{yuehua2}.
The ranges for 4.4 MeV alphas, after leaving the source coating layer, in xenon gas at different pressures, along with the energy loss rate
and energy deposition in the 0.7~$\rm{mm}$ gap before entering the drift region are simulated using the SRIM package as well.
Results are summarized in Table~\ref{yuehua.t2}. The alpha energy losses are considered when calculating the number of primary ionization electrons for the gain estimation.

\begin{table}[h]
\caption{AVERAGE RANGES AND ENERGY LOSSES OF 4.4 MEV ALPHA PARTICLES IN XENON GAS IN ROOM TEMPERATURE}
\centering
\begin{threeparttable}[h]
\begin{tabular}{ l l l l }
\toprule
{\bf Pressure} & {\bf Average Range} & {\bf $\bf dE/dx$} & {\bf Energy loss in 0.7mm}\\
& & & {\bf thick xenon} \\
{[atm]} & {[mm]} & {[MeV/mm]} & {[MeV]} \\
\midrule
1 & 15.74 & 0.21 & 0.15 \\
2 & 7.87 & 0.41 & 0.29 \\
3 & 5.25 & 0.62 & 0.44 \\
\bottomrule
\end{tabular}
\end{threeparttable}
\label{yuehua.t2}
\end{table}

\section{Results and discussion}
\label{results}
\subsection{Alpha and gamma spectrum}
\begin{figure}[htbp]
\begin{center}
\includegraphics[width=0.9\linewidth]{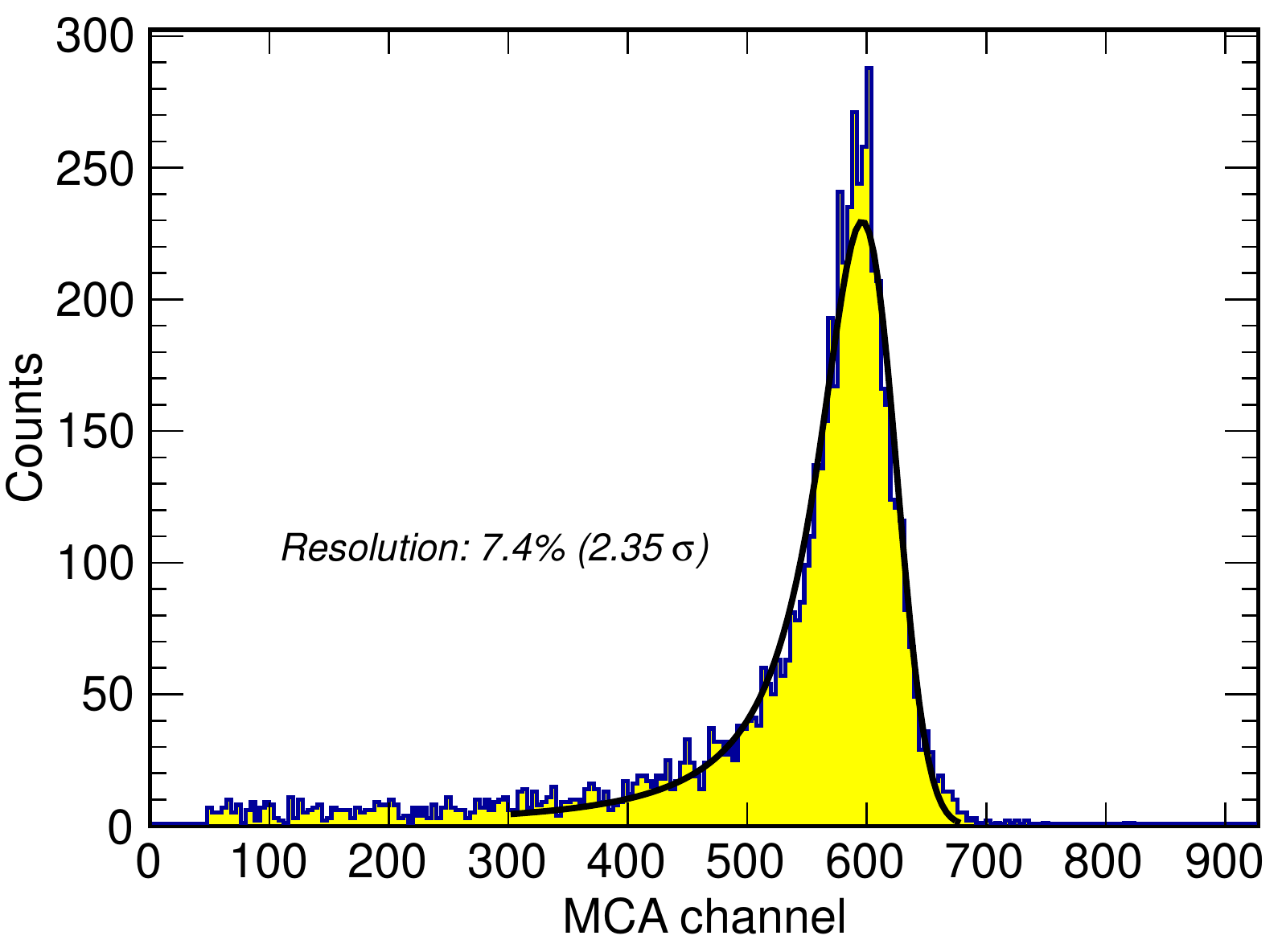}
\end{center}
\caption{An example of $^{241}$Am alpha spectrum in 1~$\rm{atm}$ xenon gas.
The voltages on the micromesh and the cathode were 430~$\rm{V}$ and 1863~$\rm{V}$ respectively.
The detected charge on the anode for each MCA channel is about 0.17~$\rm{fC}$.
The equivalent gas gain, including the electron transmission factor through the micromesh, is estimated to be 3 for this run.}
\label{yuehua3}
\end{figure}

An example of the alpha spectrum, recorded by MCA, is shown in Fig.~\ref{yuehua3}.
A long tail is presented at the lower end which is due to the source coating, as well as the ionization electrons attaching to impurities during the drift.
More details are discussed in the Sec.~\ref{psa}. In order to obtain the peak position for calculating the detected number of electrons and the gas gain,
an inverse-Landau convoluted Gaussian function is used to fit the spectrum and it is given by
\begin{equation}
\begin{split}
N(x|p_0,p_1,p_2,\sigma)=&\frac{p_2}{\sqrt{2\pi} \cdot p_0 \cdot \sigma}\times\int_{-\infty} ^\infty \! \mathcal{L}(-\frac{y+p_1}{p_0})\cdot \\
& exp(-\frac{(x+y)^2}{2{\sigma}^2}) \, \mathrm{d}y
\end{split}
\end{equation}
where $p_{0}$, $p_{1}$, $p_{2}$ are parameters of the Landau function and $\sigma$ represents the resolution for the convoluted Gaussian function.

\begin{figure}[hbt]
\begin{center}
\includegraphics[width=0.9\linewidth]{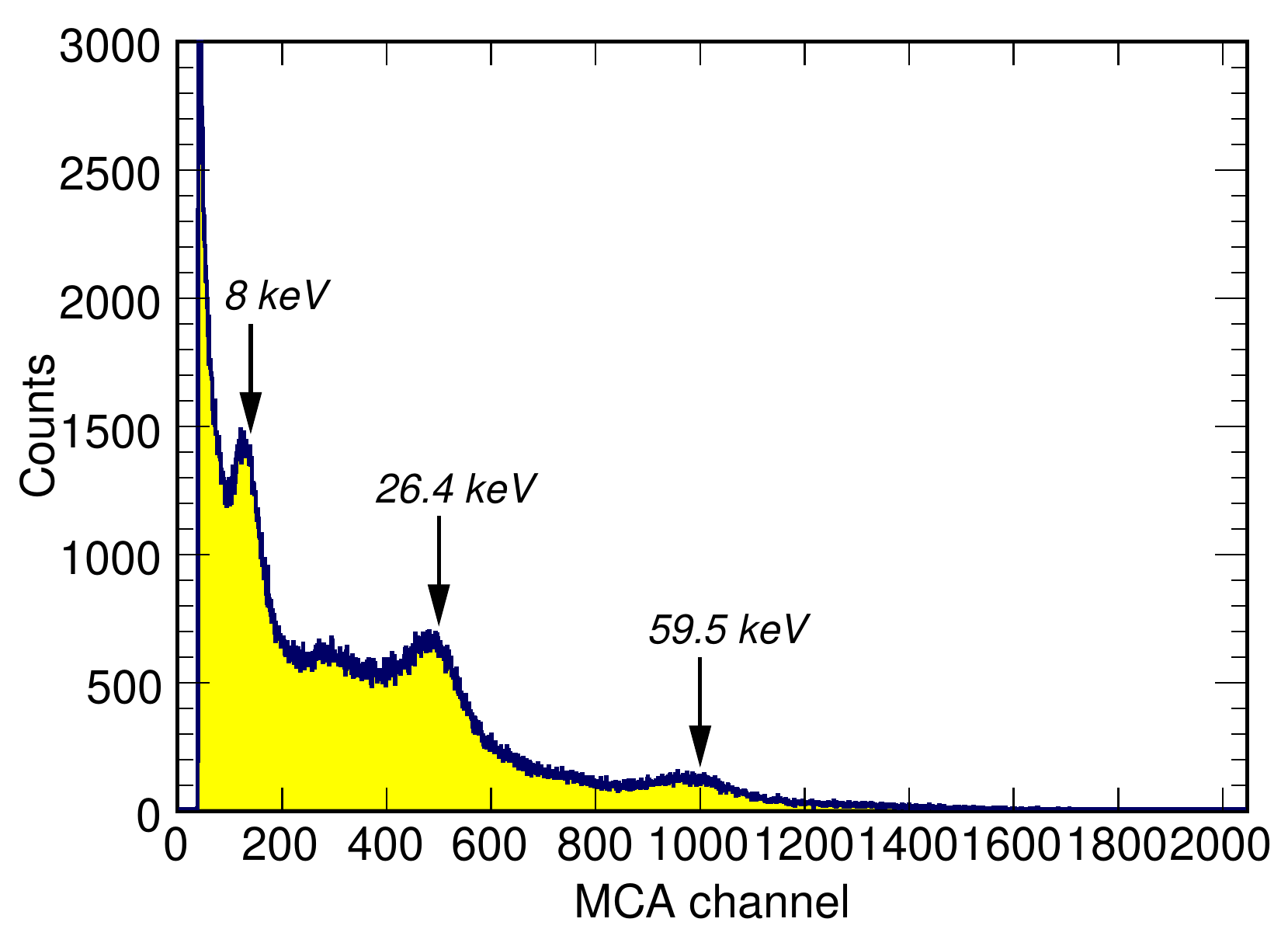}
\end{center}
\caption{A typical gamma spectrum from the $^{241}$Am source in 1~$\rm{atm}$ xenon gas that was recorded by the MCA.
The micromesh and cathode voltages were set at 550~$\rm{V}$ and 825~$\rm{V}$ respectively.
Gamma ray peaks with energies of 26.4~$\rm{keV}$, 59.5~$\rm{keV}$ and 8~$\rm{keV}$ fluorescence X-ray are clearly visible.
The detected charge on the anode for each MCA channel is about 0.17~$\rm{fC}$. The gas gain in this measurement is estimated to be 378.}
\label{yuehua4}
\end{figure}

In addition to the 5.49~$\rm{MeV}$ alpha particles, low energy gamma and Np L X-rays are emitted
by the $^{241}$Am source. Fig.~\ref{yuehua4} shows a typical gamma spectrum obtained in 1~$\rm{atm}$ xenon gas.
It should be noted that the gamma spectrum has the same charge conversion scale, about 0.17 pC per MCA channel, as for the alpha spectrum in Fig.3.
But the gamma spectrum was acquired at a much higher gain.
Gamma lines at 26.4~$\rm{keV}$, 59.5~$\rm{keV}$ and also 8~$\rm{keV}$ fluorescence X-rays from the cathode copper are observed.
The observed peak at 59.5~$\rm{keV}$ is lower than the peak at 26.4~$\rm{keV}$,
which is likely due to the smaller photoelectric absorption ratio and larger energy smearing of the 59.5~$\rm{keV}$ gamma rays.
The exponentially distributed background also pushes up the peak at 26.4~$\rm{keV}$.
Other low energy X-rays from Np L X-rays at 13.9~$\rm{keV}$, 17.7~$\rm{keV}$ and 21.0~$\rm{keV}$ also contribute to the spectrum,
but their peaks are not well resolved.

\subsection{Electron transmission and gas gain}
\begin{figure}[hbt]
\begin{center}
\includegraphics[width=0.9\linewidth]{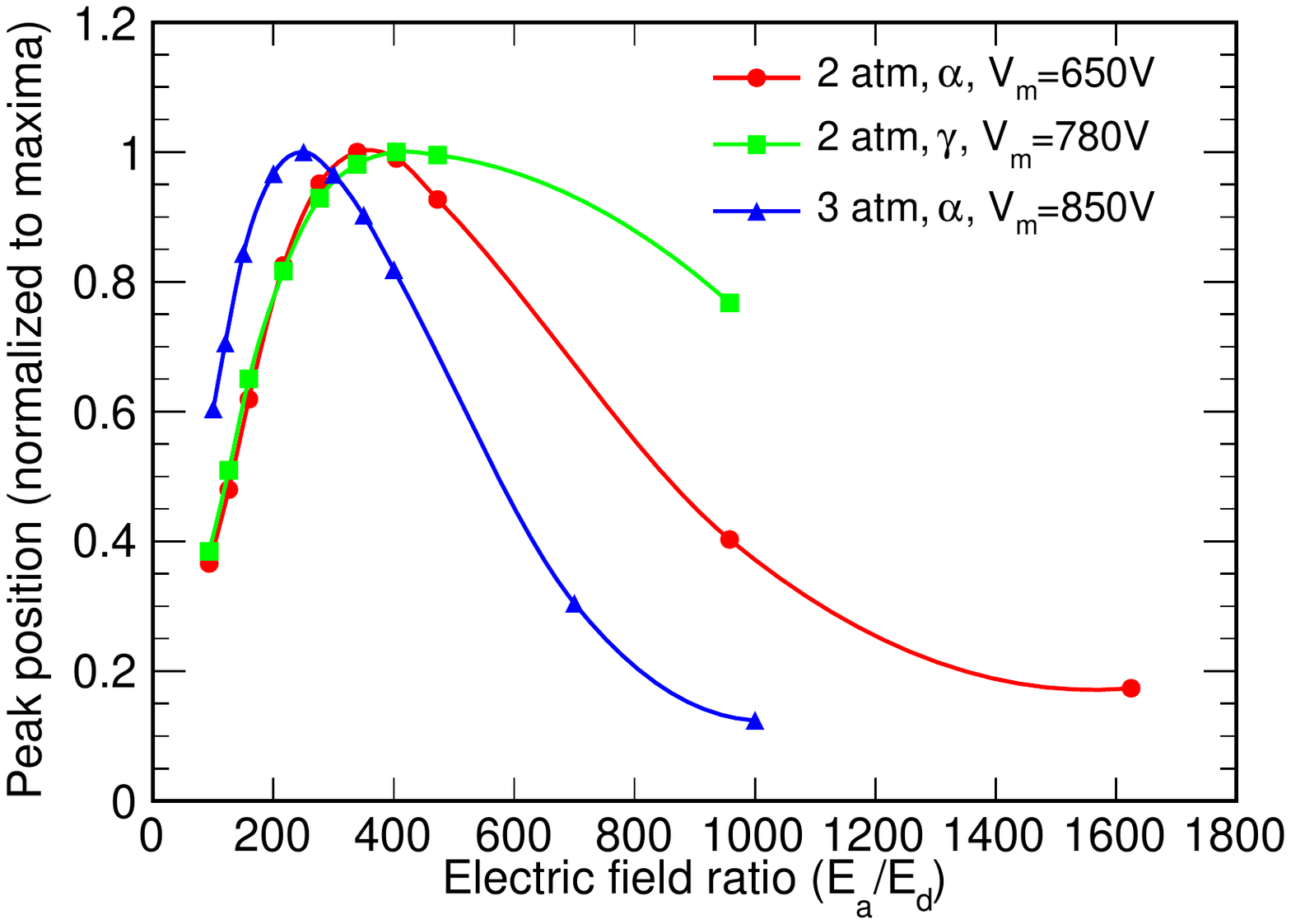}
\end{center}
\caption{Dependence of alpha and 59.5~$\rm{keV}$ gamma peak with the electric field ratio ($E_{a}/E_{d}$).
$V_{m}$ was the voltage applied on the micromesh. $E_{a}$ and $E_{d}$ stand for the electric field in the avalanche and drift gaps respectively.}
\label{yuehua5}
\end{figure}

\begin{figure}[htbp]
\begin{center}
\includegraphics[width=0.9\linewidth]{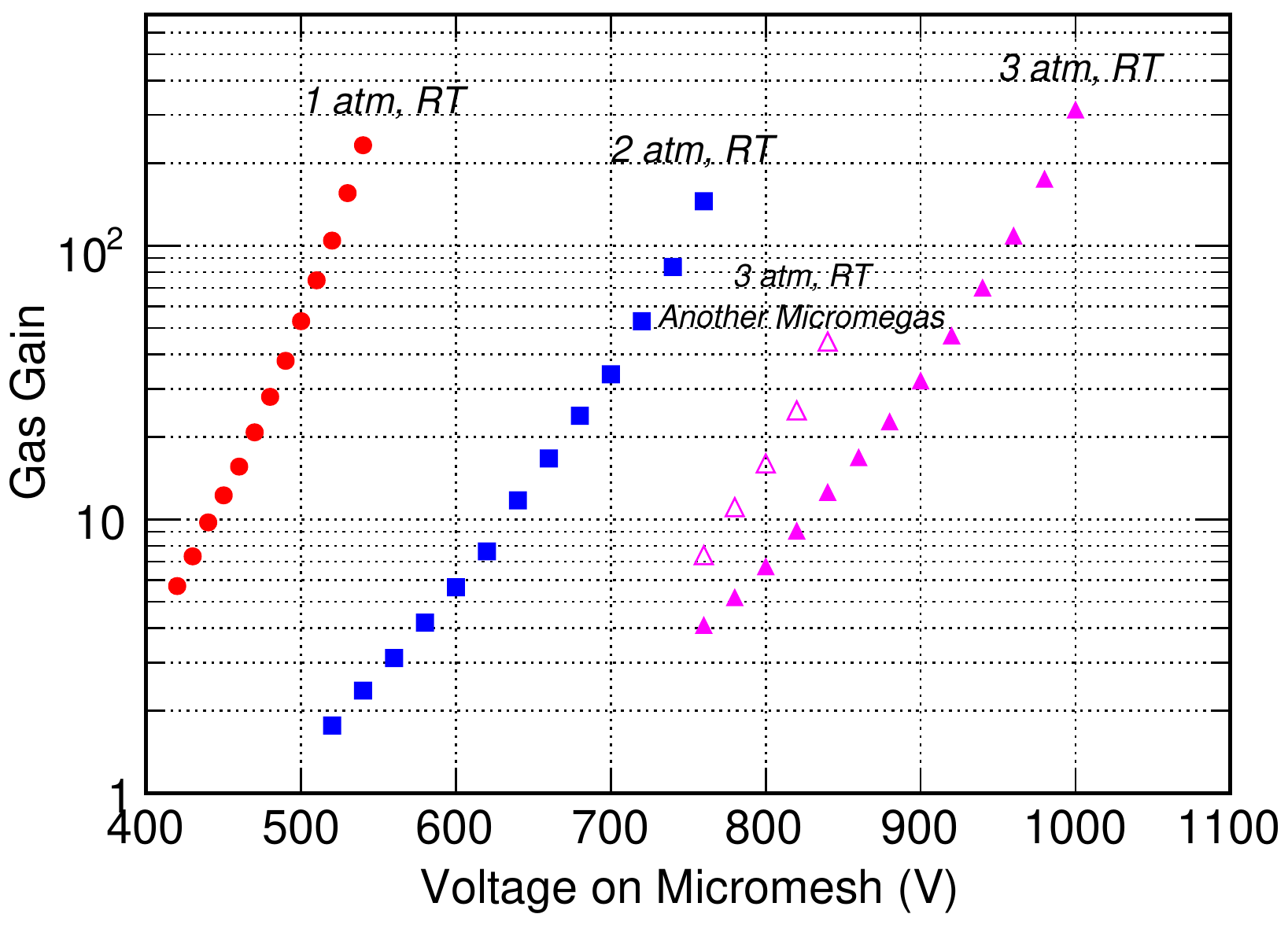}
\end{center}
\caption{The measured gas gain as a function of voltage on the micromesh. RT stands for the room temperature.}
\label{yuehua6}
\end{figure}

The transmission rate of primary electrons in the drift gap through the micromesh affects the total amount of charge collected on the anode.
In order to maximize the charge collection~\cite{Mir_2005},
we scanned the field ratios between the avalanche and drift region to reach a maximum transmission rate.
The dependence of the peak channel with the field ratio is shown in Fig.~\ref{yuehua5}, where peak channels are normalized to the maximum of each scan.
The electron transmission curves measured with alphas and 59.5~$\rm{keV}$ gamma rays agree well except at high field ratios
when ionizations from alpha particles have a much higher recombination probability. The maximum electron transmission rates are found to be at field ratios
of around 350 and 250 for 2 and 3~$\rm{atm}$ pressures respectively. These values are quite high and limit the drift fields in the detector to only a few hundred~V/cm.
The field ratio for the maximum transmission rate in 1~$\rm{atm}$ pressure xenon gas is very high,
thus was fixed to 400 to keep a drift field of at least 100~V/cm during the gain measurements.
For the gas gain estimation, we assume the electron transmission is 100\%. In other words,
the gas gain values presented in Fig.~\ref{yuehua6} is a combination of the electron transmission and the absolute gas gain.
In order to operate the Micromegas at a lower field ratio with a maximum transmission rate in the future,
the Micromegas should be made with a thinner micromesh with larger pitches.

The detector was configured with the field ratio allowing maximum charge collection efficiency in different pressures for the gas gain measurements.
The gas gain is defined as the collected number of electrons on the anode divided by the number of ionization electrons produced based on the energy deposition of alpha particles and a W value of 22~eV in xenon gas~\cite{Blum_1993}. The electronic chain with the detector connected was calibrated by injecting a known charge into the charge sensitive amplifier through a 1 pF capacitor. The corresponding charge with the registered channel in the MCA was then derived.
The results of gas gain as a function of the applied voltage on the micromesh are shown in Fig.~\ref{yuehua6}.
Gas gains exceeding 200 were achieved at pressures up to 3~$\rm{atm}$ at room temperature.
The measurement at higher gains was limited by the dynamic range of the electronics chain.
Without the limitation of the electronics chain, a maximum gain of more than 300 could be achieved by increasing the voltage by 10-30~$\rm{V}$
on the micromesh before embarking regular discharges.

We repeated the measurement with another Micromegas prototype with the same type at 3~atm xenon gas in the room temperature (Fig.~\ref{yuehua6}).
The gain for the prototype shows similar trend of increasing with an increase of the voltage on the micromesh,
but with a slightly higher gain values than those from the first prototype at the same voltage.
One reason for the gain difference between the two prototypes is the difference of the avalanche gaps produced during the micromesh attaching process.
At the time of measurement, the only charge sensitive pre-amplifier available to us is the ClearPulse 580,
which has a higher gain than the Ortec 142AH used for the first Micromegas prototype.
This limited our capability to measure the gas gain only up to 45.

\subsection{Pulse analysis and energy resolution}
\label{psa}
In these measurements, the xenon gas was not purified through any purifier. Electronegative impurities play an important role by attaching to the primary ionization electrons.
Alpha particles entering the drift gap with different incident directions produce ionized electrons with different mean drift lengths.
Generally, electrons from alpha tracks more parallel to the cathode plane arrive at the anode in a closer time period, which produce a signal with a shorter rise time.
These electrons also drift longer than the electrons from alpha tracks more perpendicular to the cathode plane. Therefore they have a larger probability to be attached to the
electronegative impurities, resulting a smaller pulse amplitude.

\begin{figure}[htbp]
\begin{center}
\includegraphics[width=0.9\linewidth]{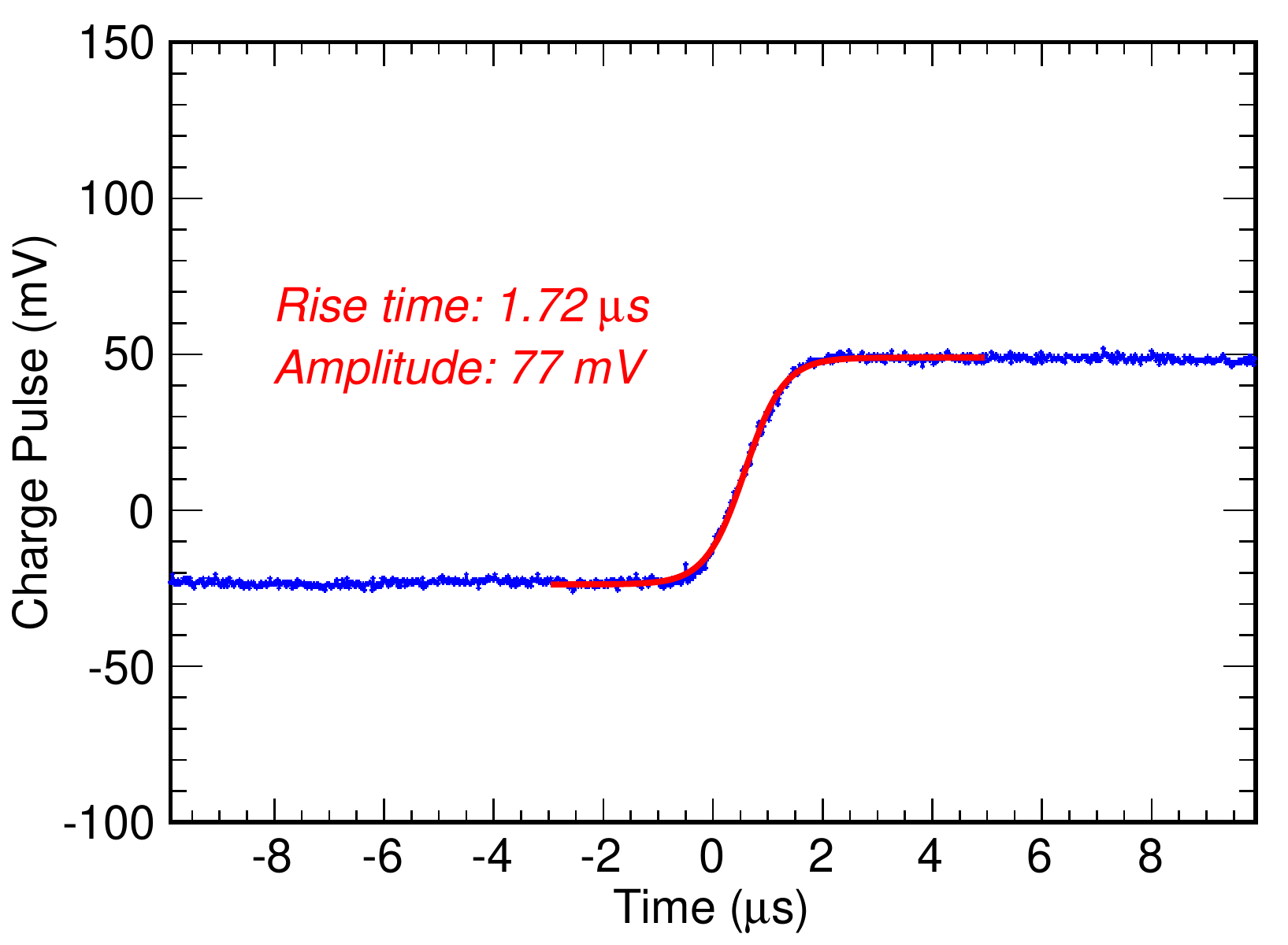}
\end{center}
\caption{A typical output pulse shape from the pre-amplifier for a detected alpha particle in 3~$\rm{atm}$ pressure xenon gas.
The cathode and mesh voltages were 1404 V and 780 V respectively.}
\label{yuehua7}
\end{figure}

\begin{figure}[htbp]
\begin{center}
\includegraphics[width=0.9\linewidth]{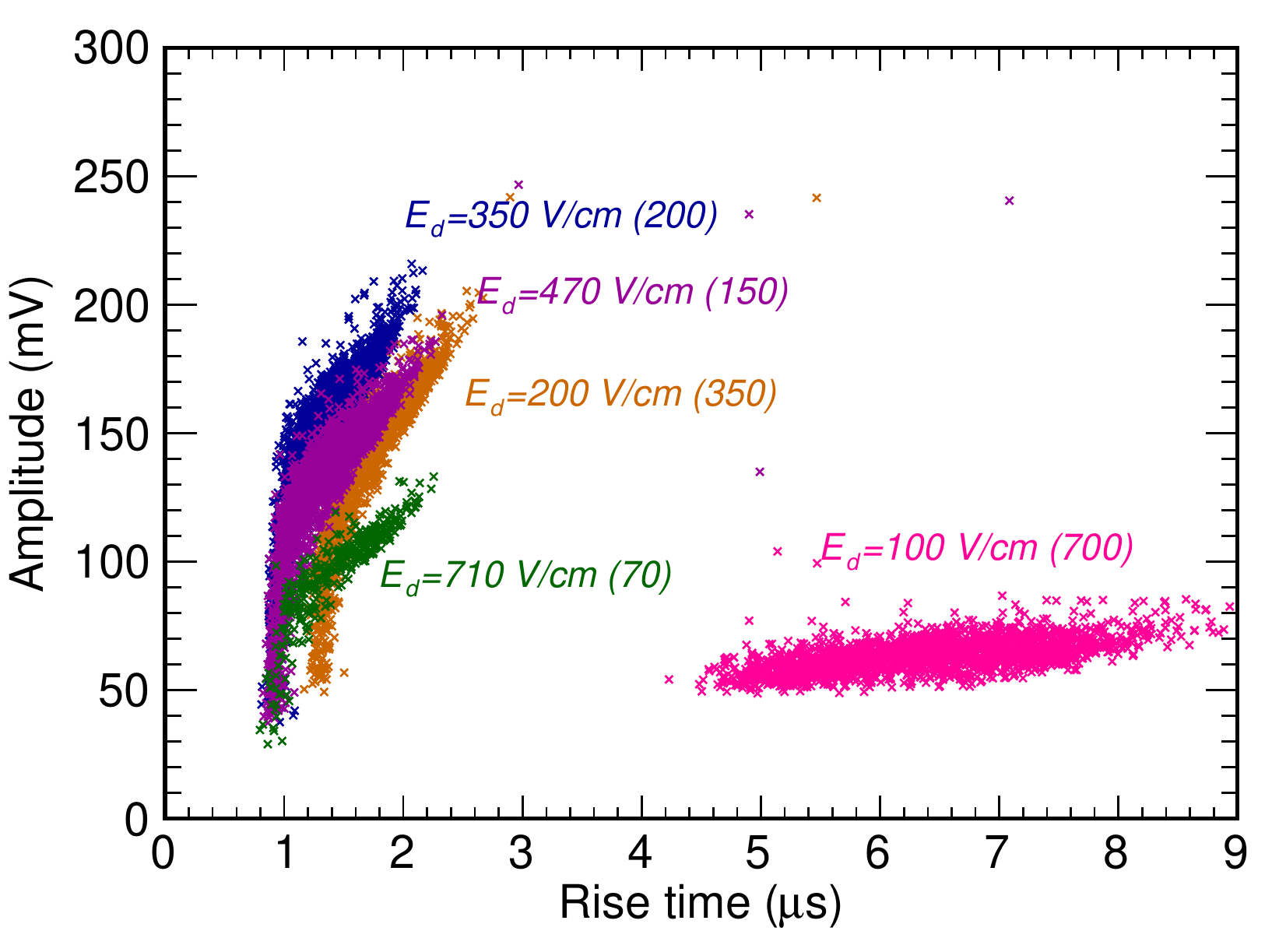}
\end{center}
\caption{The correlation of the signal rise time and the amplitude at different drift fields in 3~$\rm{atm}$ pressure xenon gas.
The voltage on the micromesh was fixed at 850~V. The field ratios ($E_{a}/E_{d}$) are shown in the parentheses in the plot.}
\label{yuehua8}
\end{figure}

To understand the significance of such an effect and to evaluate the best possible energy resolution for such a Micromegas detector,
the alpha signals after the pre-amplifier were directly recorded by the oscilloscope for off-line analysis.
A typical recorded pulse is shown in Fig.~\ref{yuehua7}. A sigmoid function given by
\begin{align}
A(t)=-&\frac{A_{tot}}{1+exp[(t-t_{half})/s]}+C
\end{align}
was used to fit the data to extract its amplitude and rise time. $A_{tot}$, $t_{half}$, $s$ and $C$ are the signal amplitude, time at half amplitude,
the sigmoid function shape parameter and the baseline offset respectively. The signal rise time is defined as the time interval between 10\% and 90\% of the total amplitude.
The correlation between the signal rise time and the amplitude for different drift fields in 3~$\rm{atm}$ pressure xenon are shown in Fig.~\ref{yuehua8}.
As expected, the amplitude increases when the drift field was increased from 100~V/cm to 350~V/cm.
But it starts to decrease when the drift field is higher than 350 V/cm, which is due to the smaller electron transmission rate at a lower field ratio.

\begin{figure}[htbp]
\begin{center}
\includegraphics[width=1.0\linewidth]{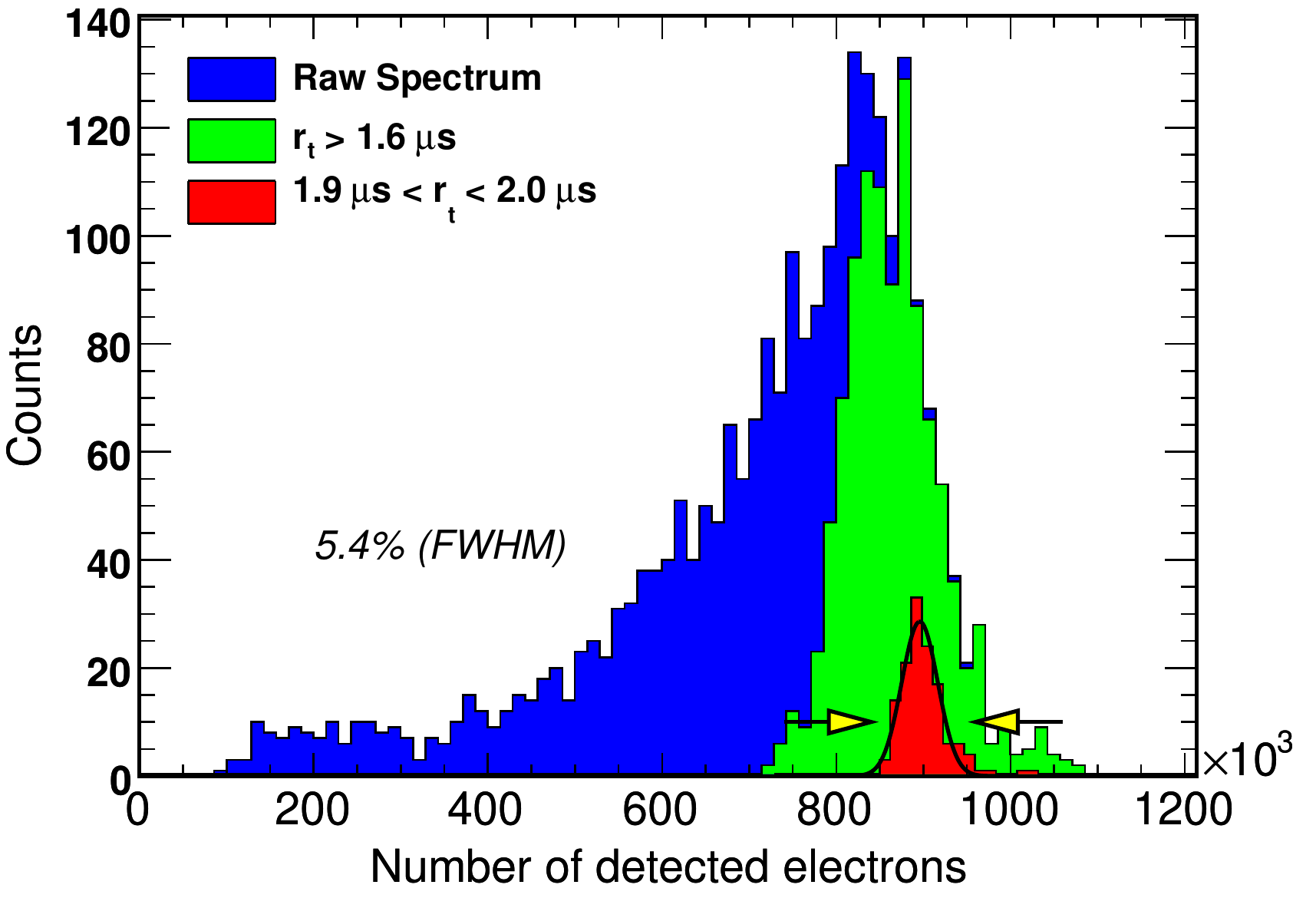}
\end{center}
\caption{Typical distribution of the alpha signal amplitudes in 298 K and 3~ atm xenon gas before and after the rise time selection.
The voltages on the micromesh and the cathode were 780 V and 1404 V respectively.
The number of detected electrons is about $9 \times 10^5$, corresponding to a gas gain of 5. $\rm{r_{t}}$ stands for the rise time.}
\label{yuehua9}
\end{figure}

In order to understand the influence of ionization attachment on the energy resolution, cuts based on the rise time are applied
to the amplitude distribution as shown in Fig.~\ref{yuehua9}.
The selection of the events with longer rise time, corresponding to alpha particles with tracks more perpendicular to the micromesh,
removes the long tail at the lower end of the spectrum.
A 100~$\rm{ns}$ cut on the rise time gives an energy resolution of 5.4\% (FWHM).
This value is almost the best we could achieve with the Micromegas in the current setup.
To achieve a better energy resolution closing to the statistical limit from the Fano factor and multiplication variance,
we need to improve the Micromegas fabrication technique and to purify the xenon gas to suppress the ionization attachment to impurities.

\begin{figure}[hbt]
\begin{center}
\includegraphics[width=0.95\linewidth]{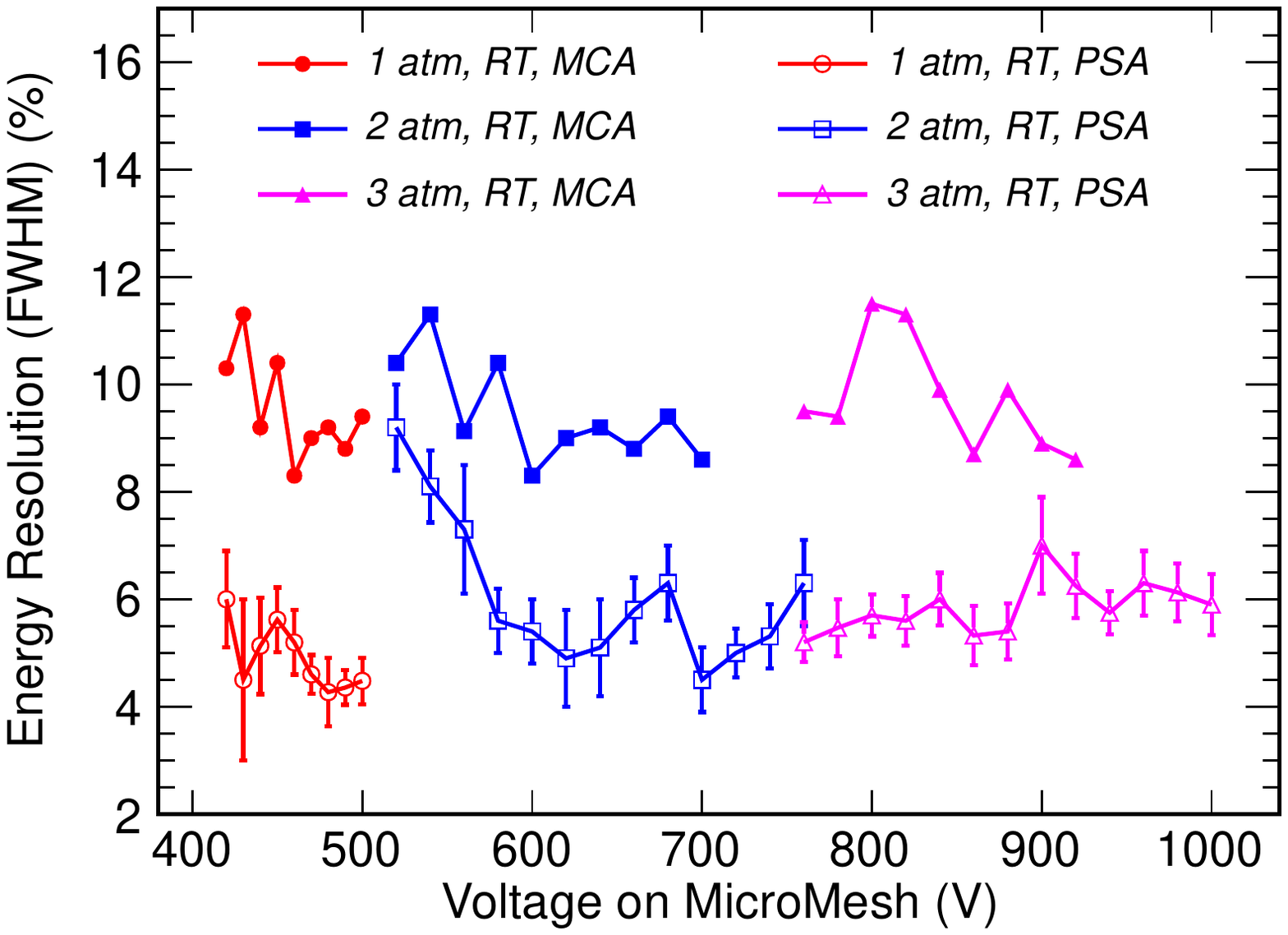}
\end{center}
\caption{The energy resolution for alphas as a function of voltage on the micromesh in different pressures and temperature of xenon gas.
Solid symbols: extracted from fitting spectra that was recorded by MCA with inverse-Landau convoluted Gaussian function.
Open symbols: extract from analyzing the pulse shape that recorded by oscilloscope.
PSA stands for the pulse shape analysis method.
RT stands for the room temperature.}
\label{yuehua10}
\end{figure}

The resolution at different voltages in different pressures are plotted in Fig.~\ref{yuehua10} (open symbols)
together with those obtained from the function fitting the MCA recorded spectra (solid symbols).
The achievable resolutions are worse than those obtained with a microbulk Micromegas detector~\cite{Dafni_2009,Tomos_2011}
and it is very likely due to the source coating which increases the energy deposition fluctuations,
as well as the non-uniform gas gap in a high field operation. Unlike the microbulk Micromegas, the micromesh here is supported only at the edges.
It deforms at a strong field between the micromesh and the anode pad.

\section{Conclusion}
\label{concl}

We have studied the characteristics of a Micromegas prototype fabricated by the thermo-bond method
in different pressures of xenon gas at room temperature. From our measurement, low energy gamma rays down to 8 keV can be
observed. Gas gains above 200 are achieved in pure xenon gas up to 3~$\rm{atm}$ pressure with the alpha particles from a $^{241}$Am source.
The resolution is better than 9\% (FWHM) using the MCA recorded spectrum and is further improved to 5\% by using a pulse shape analysis.
It should be noted that these obtained results also include fluctuations due to energy loss in the surface coating of Americium source.

Our investigations of the Micromegas operating at room temperature show its potential application for low energy event detection in xenon gas.
In order to produce a device to be used in dark matter or neutrinoless double beta decay searches,
further optimization of the fabrication technique, such as using a thinner avalanche gap from a thinner micromesh,
needs to be done in order to improve the gain and energy resolution, especially in the low temperature environment.

An attempt was made to measure the gas gain of the Micromegas at the cryogenic temperature in xenon gas to investigate its potential application in a two-phase xenon detector.
However the cryogenic system was not optimized enough to give a systematic result at this time.
We will report the results of operating the Micromegas in a two-phase xenon detector in the future.

\section*{Acknowledgement}
The authors would like to thank Prof. Ioannis Giomataris and Dr. Joerg Wotschack for discussions of Micromegas development during their visits to China.

\ifCLASSOPTIONpeerreview
\begin{center} \bfseries EDICS Category: 3-BBND \end{center}
\fi
%
\IEEEpeerreviewmaketitle

\end{document}